ARTÍCULOS

# Mapas científicos de la Revista General de Información y Documentación (2005-2022)

**Carmen Gálvez**
Profesora, Dept. Información y Comunicación, Universidad de Granada (España) ✉ iD

**ES Resumen.** Se presenta un estudio de la Revista General de Información y Documentación (RGID), desde 2005 hasta 2022, con el objetivo de calificar la estructura de su campo de investigación y valorar la trayectoria de las áreas temáticas tratadas. La metodología fue el análisis de co-palabras, la construcción de redes bibliométricas y la creación de mapas científicos. Se extrajeron 514 documentos de la base de datos *Web of Science* (WoS). Se seleccionaron las palabras-clave asignadas por los autores de los documentos y se dividieron en tres subperiodos: 2005-2010, 2011-2016 y 2017-2022. En los resultados se obtuvieron 1701 palabras-claves de autor y 37 redes bibliométricas. En el periodo 2005-2010, la estructura del campo de investigación se distribuyó en el mapa científico con muy pocos temas centrales y especializados, considerándose una organización incipiente y poco desarrollada. En el periodo 2011-2016, la estructura del campo de investigación se distribuyó en el mapa científico con un número más variado de temas centrales y especializados, pero todavía insuficiente, considerándose una organización en vías de desarrollo. En el periodo 2017-2022, la estructura del campo de investigación se mostró en el mapa con todo tipo de familia de temas (centrales, especializados, transversales, emergentes o en desaparición), valorándose como una organización dinámica, compleja y heterogénea. En cuanto a la evolución de las áreas temáticas, el mapa presentó un sólido progreso entre los dos últimos periodos. La morfología del campo temático tratado en RGID, se esquematizó en tres fases: inicio (2005-2010), vías de desarrollo (2011-2016) y consolidación (2017-2022).
**Palabras clave.** Revista general de información y documentación, RGID, bibliometría, análisis de co-palabras, mapas científicos, visualización de la información, análisis de revistas.

## ENG Science mapping of the Revista General de Información y Documentación (2005-2022)

**ENG Abstract.** A study of the Revista General de Información y Documentación (RGID)General Magazine of Information and Documentation (RGID), from 2005 to 2022. The objective is aimed at qualifying the structure of the research field and assessing the trajectory of the thematic areas covered. Applying as methodology the analysis of co-words, the construction of bibliometric networks and the creation of scientific maps. 514 documents are extracted from the Web of Science (WoS) database. The keywords assigned by the authors of the documents are selected and divided into three subperiods: 2005-2010, 2011-2016 and 2017-2022. In the results, 1701 author's keywords and 37 bibliometric networks are obtained. In the period 2005-2010, the structure of the research field is represented on the scientific map with very few central and specialized topics, considering an initial and underdeveloped organization. In the period 2011-2016, the structure of the research field is distributed on the scientific map with a more varied number of central and specialized topics, but still insufficient, considering an organization in the process of development. In the period 2017-2022, the structure of the research field is shown on the map with all kinds of family of topics (central, specialized, transversal, emerging or disappearing), being valued as a dynamic, complex and heterogeneous organization. Regarding the evolution of the thematic areas, the map shows solid progress between the last two periods. The morphology of the thematic field treated in RGID is outlined in three phases: foundation (2005-2010), process of development (2011-2016) and consolidation (2017-2022).

**Sumario.** 1. Introducción. 2. Metodología. 3. Resultados. 4. Discusión. 5. Conclusiones. 6. Referencias bibliográficas.

## 1. Introducción

Las revistas científicas forman una parte central del proceso de comunicación académica y son pieza integral de la propia investigación científica (Ware y Mabe, 2015). Al mismo tiempo, este tipo de recurso se puede utilizar como fuente de información para otros objetivos, tales como conocer cuáles son los autores, o instituciones, con una mayor productividad e impacto o los frentes de investigación en un campo de conocimiento (Abadal, 2017). En el caso particular de la información científica, para lograr este propósito es necesario aplicar una serie de metodologías que permitan procesar la información y transformarla en conocimiento, con especial relevancia del método bibliométrico (Van Raan, 2005; Glänzel y Moed, 2013).

Los estudios bibliométricos tienen como finalidad, por un lado, el tratamiento y análisis cuantitativo de la bibliografía científica y, por el otro, el estudio de la estructura y la dinámica social que la producen y utilizan (López-Piñero, 1972). La bibliometría aplica diversos indicadores para explorar un campo de investigación a partir de datos bibliográficos (Noyons, Moed y Van Raan, 1999; Bordons y Zulueta, 1999; Van Raan, 2004; Van Raan, 2005). Básicamente se distinguen dos tipos de parámetros. Indicadores unidimensionales (basados en técnicas estadísticas univariables, dedicados a analizar o medir una única característica de los documentos seleccionados, sin tener en cuenta ningún vínculo que pudiera haber entre ellos) e indicadores multidimensionales, o indicadores relacionales, (basados en técnicas estadísticas multivariantes, dedicados a medir de forma simultánea diferentes unidades de análisis en los documentos seleccionados). Dentro de los indicadores relacionales se sitúa la construcción de los mapas científicos (también denominados mapas bibliométricos, cienciogramas o diagramas estratégicos), dirigidos básicamente a analizar diferentes estructuras (sociales, intelectuales o conceptuales) en la literatura científica (Börner *et al*., 2003; Noyons, Moed y Luwel, 1999; Small, 1997; Morris y Van der Veer Martens, 2008).

En el caso específico de la evaluación de revistas académicas, la bibliometría tiene un papel importante encuadrándose en el contexto científico de los denominados «estudios sociales de la ciencia» (Bordons y Zulueta, 1999). Dentro del área de conocimiento de las Ciencias de la Documentación, se encuentran diversos trabajos dirigidos al análisis de revistas aplicando indicadores bibliométricos (DeHart, 1992; Lipetz, 1999; Koehler, 2001; Bonnevie-Nebelong, 2003); en relación a los estudios sobre revistas españolas también han destacado diversos trabajos (Pérez-Álvarez-Ossorio, 1997; Giménez-Toledo y Román-Román; 2000; López-López *et al.,* 2001; Jiménez-Hidalgo, 2007; Ollé Castellà y Porrás, 2008; Guallar *et al*., 2017; López-Robles *et al*., 2019; Cascón-Katchadourian *et al*., 2020; Guallar *et al*., 2020).

El objetivo de esta investigación fue el estudio de la publicación académica Revista General de Información y Documentación (RGID), en el marco temporal de los últimos 18 años (2005-2022), que son los que lleva la revista siendo indexada en la *Web of Science* (WoS), para responder a las siguientes preguntas de investigación:

− ¿Cuál es la estructura conceptual y temática de la revista?
− ¿Cómo se califica y caracteriza la estructura del campo de investigación de la revista?
− ¿Cómo han evolucionado las áreas temáticas de la revista?

## 2. Metodología

El método para la detección de la estructura conceptual y temática de la RGID fue el análisis de palabras asociada, este procedimiento se ha utilizado en diferentes estudios, con gran robustez y un amplio recorrido en investigación (Callon, Rip y Law, 1986; Callon, Courtial y Laville, 1991; Cobo *et al*., 2011; Leydesdorff y Welbers, 2011; Liu *et al.,* 2012; Raeeszadeh *et al.,* 2018). El análisis de las palabras asociadas, o análisis de co-palabras, se dirige a la creación de mapas científicos (o temáticos) donde se identifiquen los focos o centros de interés de un campo. Estos focos son asimilables a las áreas temáticas que componen un área estudiada. El análisis de las palabras asociadas se realiza a partir de un conjunto de documentos representativos de la producción de un área y la construcción de redes bibliométricas y mapas científicos (o diagramas estratégicos). Esta técnica parte del principio según el cual una especialidad de investigación puede ser identificada por su propio vocabulario. A partir del cómputo de las apariciones conjuntas de palabras, representadas como relaciones, se construyen redes de términos que muestran la estructura temática del campo de investigación analizado.

En este trabajo, el procedimiento seguido para la construcción de las redes bibliométricas y los mapas científicos se desarrolló en varias etapas: 1) recopilación de datos; 2) selección de las unidades de análisis; 3) pre-procesamiento de las unidades de análisis; 4) creación de las redes bibliométricas; 5) construcción de los mapas científicos; 6) visualización de los mapas científicos; 7) construcción del mapa de evolución de las áreas temáticas; y 8) interpretación de los mapas científicos. Para el procesamiento estadístico y descriptivo de los datos se empleó el paquete RStudio Bibliometrix versión 4.1.0 y la aplicación Biblioshiny (interfaz web de Bibliometrix) (Aria y Cuccurullo, 2017), se trata de una herramienta de código abierto, desarrollada en R, que incluye los principales indicadores del análisis bibliométrico. Para la construcción de los mapas científicos se utilizó la herramienta informática SciMAT (*Science Mapping Analysis Software Tool*) (Cobo *et al*., 2012).

### 2.1. Recopilación de datos

La fuente de información utilizada fue la base de datos Web of Science (WoS) (Clarivate Analytics, Philadelphia, PA, USA). La elección de WoS se debió a que proporciona numerosas herramientas de análisis para procesar los datos y ofrece información de investigación altamente precisa. Dentro de la colección principal de WoS, los datos se obtuvieron de las bases de datos Science Citation Index Expanded (SCI-EXPANDED), Social Sciences Citation Index (SSCI) y Arts & Humanities Citation Index (A&HCI) y Emerging Sources Citation Index (ESCI). La estrategia de búsqueda consistió en seleccionar el campo título de publicación (Source): SO = (REVISTA GENERAL DE INFORMACION Y DOCUMENTACION).

### 2.2. Selección de las unidades de análisis

Como se pretendió analizar la estructura conceptual y temática de la publicación RGID, se seleccionaron las palabras-clave asignadas por los autores de los documentos. En la base de datos WoS, los registros incluyen dos tipos de palabras-clave: palabras-clave de autor (*Author's keywords*), proporcionadas por los propios autores, y palabras-clave (*Keywords Plus*), que son términos del índice generados automáticamente a partir de los títulos de los artículos citados. Para el análisis bibliométrico, las palabras-clave extraídas de forma automática son menos específicas que las palabras-clave aportadas por los propios autores (Zhang *et al.*, 2016), por esta razón se decidió seleccionar las palabras-clave de autor.

### 2.3. Pre-procesamiento de las unidades de análisis

La construcción de los mapas científicos no se realizó directamente sobre los datos extraídos de la base de datos WoS, porque los posibles errores y duplicados afectarían a la calidad de los resultados. Para ello, se realizó una fase pre-procesamiento de las unidades de análisis, que incluyó:

- Unificación de elementos duplicados, similares o escritos erróneamente, además se unificaron los plurales y singulares de las palabras-clave.
- División temporal de los datos en tres subperiodos, con el propósito de analizar los temas de investigación en cada etapa y la evolución longitudinal de los temas de investigación tratados en la revista: 2005-2010, 2011-2016 y 2017-2022.
- Aplicación de medidas de reducción de datos para disminuir el tamaño de la muestra para y los resultados fueran más comprensibles. Se seleccionaron las palabras-clave de autor que aparecieran con una frecuencia mínima de 2.

### 2.4. Construcción de las redes bibliométricas

A partir de las relaciones de co-ocurrencia de dos palabras-clave se crearon redes bibliométricas. Las relaciones de co-ocurrencia de palabras-clave se representaron como redes de términos, en los que los nodos representaron las palabras y las relaciones entre ellas representaron los enlaces (la fuerza de los enlaces dependió del número de apariciones conjuntas en un mismo documento, si se dieran muchas co-ocurrencias sobre una misma palabra, darían lugar a una alianza estratégica entre documentos que se asociaría con un tema de investigación).

Una vez creadas las redes de palabras-clave se precisó calcular la intensidad de las asociaciones entre palabras, para ello fue necesario normalizarla los enlaces (Van Eck y Waltman, 2009). Se dice que dos palabras-clave, $i$ y $j$, co-ocurren si aparecen juntas en la descripción de un documento. Sin embargo, el simple conteo de co-ocurrencias no es un buen método para evaluar los enlaces entre dos palabras. Fue necesario adoptar índices estadísticos para normalizar los valores de co-ocurrencia (Salton y McGill, 1983; Callon, Courtial y Laville, 1991; Van Raan, 2004). En este trabajo, se aplicó el índice de equivalencia (Callon, Courtial y Laville, 1991) como medida de similitud, que se define como: sea $c_i$ el número de ocurrencias de la palabra $i$ (es decir, el número de veces que se utiliza esa palabra para indexar un documento) y $c_j$ el número de ocurrencias de la palabra $j$ (es decir, el número de veces que se utiliza esa palabra para indexar un documento), sea $c_{ij}$ el número de co-ocurrencias de las palabras $i$ y $j$ (esto es, el número de documentos que están descritos por ambas palabras), el índice de equivalencia se calculó como: ($e_{ij}$) sería $e_{ij} = c^2_{ij}/c_i \times c_j$. La aplicación del índice de equivalencia estableció un peso a cada palabra-clave, equivalente a su importancia en el conjunto de documentos analizados. El cálculo de todos los coeficientes entre todos los pares de palabras posibles generó un número excesivo de enlaces (demasiados para poder representarlos gráficamente). Por ello, en la etapa siguiente, se aplicaron algoritmos de agrupamiento, para que las redes de palabras-clave pudieran ser más fácilmente visualizadas en mapas científicos.

### 2.5. Construcción de los mapas científicos

La creación de los mapas científicos fue la etapa principal de la metodología, se realizó a través de dos procesos (Börner, Chen y Boyack, 2003; Cobo *et al.*, 2012):

1. Aplicación de técnicas de agrupamiento (clasificación o clustering). Dentro de los procedimientos de agrupamiento de palabras-clave se utilizó el algoritmo del centro simple (Coulter, Monarch y Konda, 1998), que permitió crear redes de palabras-clave vinculadas entre sí y que se correspondieron con centros de interés, temas o problemas de investigación temática. El algoritmo usa diferentes parámetros, como umbral mínimo de frecuencia y de co-ocurrencia. De esta forma sólo las parejas de palabras-clave que superen ciertos umbrales serán consideradas como enlaces en la construcción de las redes bibliométricas. Además, el algoritmo tiene dos parámetros para limitar el tamaño de las redes detectadas. Los cuatro parámetros utilizados fueron: 2, como umbral mínimo de frecuencias, 10 tamaño máximo de los nodos de la red y 1 tamaño mínimo de los nodos de la red. La aplicación de algoritmos de agrupamiento a las redes bibliométricas se dirige a la creación de subredes con nodos fuertemente enlazados entre sí, pero poco enlazados con el resto de la red. Las técnicas de agrupamiento dividieron el conjunto de palabras-clave en subconjuntos, los cuales cumplieron la condición de tener una gran cohesión interna (es decir, los subgrupos formados deben tener una gran similitud entre sí y se deben diferenciar del resto de los grupos de la red bibliométrica global).
2. Aplicación de dos medidas para representar las redes bibliométricas detectadas (Callon, Courtial y Laville, 1991). Primera, medida de centralidad para calcular el grado de interacción de una red con respecto a otras redes, se define como como $c = 10 * \sum e_{kh}$, siendo k una palabra-clave perteneciente al tema, y h una palabra clave que pertenece a otros temas (esta medida se puede interpretar como el grado de importancia y relevancia de un tema en el desarrollo global de campo científico analizado, o como el grado de cohesión externa del tema). Segunda, medida de densidad para calcular la fuerza interna o densidad de todos los enlaces entre palabras-clave que describen el tema, una red, puede definirse como $d = 100 (\sum e_{ij}/w)$, donde i y j son palabras-clave que pertenecen al tema, y w el número de palabras-clave en el tema (esta medida se puede interpretar como el grado de desarrollo y especialización de un tema en el desarrollo global de campo científico analizado).

### 2.6. Visualización de los mapas científicos

Los conjuntos de grupos de palabras-clave interconectadas (grupos temáticos o temas), se visualizaron en forma de diagramas estratégicos (o mapas científicos). Los diagramas estratégicos constituyen visualizaciones bidimensionales de las redes de palabras-clave en cuatro cuadrantes, de acuerdo a sus valores de centralidad (grado de relevancia) y de densidad (grado de desarrollo) (Callon, Courtial y Laville, 1991). La posición estratégica de las redes bibliométricas, en los respectivos cuadrantes, proporcionó una lectura comprensible del grado de importancia y desarrollo de los temas que la componen. En los diagramas estratégicos los temas se clasificaron en cuatro grupos (Figura 1):

a) Temas centrales y desarrollados (cuadrante 1). Se encontraron los temas clasificados como temas motores, que presentaron una centralidad fuerte y una alta densidad. Configuraron los temas importantes y constituyeron el centro del campo. Los agregados de palabras-clave que lo formaron presentaron intensas relaciones con alto grado de desarrollo e integración.
b) Temas centrales y no desarrollados (cuadrante 2). Se situaron los temas clasificados como básicos, pero no desarrollados. Los clústeres de palabras-clave que lo formaron estuvieron ampliamente conectados a otras agrupaciones, pero la densidad de sus relaciones internas fue relativamente débil. Representaron los temas importantes para el desarrollo del campo (susceptibles de convertirse en centrales y desarrollados, y, por tanto, desplazarse al cuadrante 1).
c) Temas periféricos y desarrollados (cuadrante 3). Se encontraron los temas especializados y desarrollados del campo. La fuerte intensidad de sus relaciones internas (gran densidad) hizo que se correspondieran con problemáticas de investigación cuyo estudio estuvo bien desarrollado, pero cuyos enlaces externos fueron débiles. Agrupa temas que han podido ser centrales pero que se han convertido en especializados debido a que interactúan débilmente con respecto a otras subredes.
d) Temas periféricos y poco desarrollados (cuadrante 4). Se ubicaron los temas con una centralidad y densidad baja, es decir, tanto temas emergentes como temas marginales (susceptibles de desaparición).

Según la lectura que se haga de la distribución de los temas, se consideraron tres tipos de agregados (Callon, Courtial y Laville, 1991):

− Categoría 1: los temas se distribuyen alrededor de la primera bisectriz (cuadrante 1 - cuadrante 4), esto indica que la red se organiza en torno a un núcleo de temas bien desarrollados y que están en contacto con un conjunto de temas poco desarrollados y periféricos.
− Categoría 2: la distribución de temas se realiza en torno a la segunda bisectriz (cuadrante 2 - cuadrante 3), revelando que la red está en vía de estructuración, ya que hay pocos temas motores, la mayor parte de ellos se distribuyen entre temas especializados y temas transversales.
− Categoría 3: la distribución de temas se sitúa en los diferentes cuadrantes, señala que la red está bien estructurada, es compleja y valiosa. Presenta todas las familias de agregados (motores, básico, centrales, especializados, emergentes, marginales o con diversos grados de desarrollo), una estructuración de este tipo es indicativa de una excelente dinámica del campo de estudio.

Además, las visualizaciones se enriquecieron añadiendo una tercera dimensión a los elementos representados, como fue el número de documentos asociados al tema. De este modo, los temas se mapearon como esferas de redes bibliométricas (también denominadas agregados o redes temáticas), en las que el volumen fue proporcional al número de documentos vinculados a ese tema. Por último, cada red temática se etiquetó con el nombre de la palabra-clave más central, y el grosor de los enlaces (o las líneas entre dos palabras-clave) fue proporcional al índice de equivalencia.

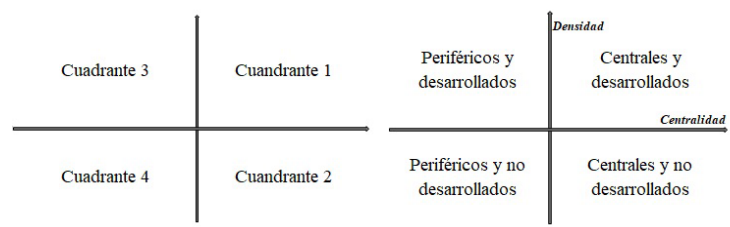

(figura.1) Distribución de los temas en los diagramas estratégicos. Fuente: Elaboración propia.

### 2.7. Construcción del mapa de evolución de las áreas temáticas

Los diagramas estratégicos mostraron la descripción estática de las redes bibliométricas de palabras-clave. Para trazar la dinámica del campo analizado a lo largo del tiempo, fue necesario presentar, primero, la evolución de las palabras-clave y, segundo, descubrir los nexos conceptuales existentes entre los temas de investigación a través de los diferentes periodos analizados.

Para analizar la evolución de las palabras-clave, se midió el grado de solapamiento entre los términos de dos periodos consecutivos aplicando el índice de inclusión (Small, 1977). Con este índice, se calculó el número de palabras-clave compartidas entre dos periodos.

Para descubrir los nexos conceptuales entre los temas, se dividió el corpus de documentos en los tres intervalos de tiempo previamente definidos. La herramienta SciMAT (Cobo *et al*., 2012) permitió crear el mapa de la evolución longitudinal de un campo, según el siguiente planteamiento: sea $T^t$ el conjunto de los temas detectados en el periodo de tiempo $t$, donde $U \in T^t$ representa cada uno de los temas detectados en el periodo $t$. Sea $V \in T^{t+1}$ el conjunto de los temas detectados en el siguiente periodo de tiempo $t + 1$. Se dice que hay una evolución temática desde el tema *U* al tema *V* si las redes temáticas de ambos temas comparten al menos una palabra-clave. De este modo, *V* puede considerarse como un tema que ha evolucionado de *U*. Las palabras-clave que pertenezcan tanto a *U* como a *V*, se considerarán como nexos conceptuales o temáticos de la evolución. De esta forma, el mapa científico de evolución se construye enlazando temas del periodo $T^t$ con temas del periodo $T^{t+1}$ por medio de nexos conceptuales entre ellos. Los nexos conceptuales se detectaron utilizando el denominado índice de inclusión (Rip y Courtial, 1984; Sternitzke y Bergmann, 2009), que mide la similitud de palabras-clave entre los diferentes periodos. Por último, las áreas temáticas se visualizaron como un grafo bipartito (o mapa de evolución) que conectaron, a través de nexos conceptuales, elementos de los dos diferentes periodos.

### 2.8. Interpretación de los mapas científicos

La interpretación de los mapas científicos no fue sencilla porque fueron muchas las medidas y procedimientos implicados. Además de los algoritmos para generar los agregados, se aplicaron diversos índices cuantificados (equivalencia, centralidad, densidad e inclusión), que hubo que definir previamente para medir la intensidad de las apariciones conjuntas y conseguir representaciones simplificadas de las redes temáticas resultantes. A continuación, hubo que interpretar distribución de las redes en los cuatro cuadrantes de los diagramas estratégicos y calificar la estructura del campo de investigación de RGID. De la misma forma, hubo que explicar la evolución de las áreas temática en el mapa longitudinal.

## 3. Resultados

Con los criterios de búsqueda seleccionados, se obtuvo un total de 514 publicaciones en la base de datos WoS. Sus características específicas se presentaron en la Tabla 1. La distribución de publicaciones por años en RGID se ha mantenido relativamente estable en el periodo completo analizado (2005-2022) (Figura 2).

Tabla 1 Información de los datos obtenidos de *Web of Science* (WoS)..

| DESCRIPCIÓN | RESULTADOS |
|---|---|
| INFORMACIÓN PRINCIPAL SOBRE LOS DATOS | |
| Período | 2005-2022 |
| Fuentes (Revistas, Libros, etc) | 1 |
| Documentos | 514 |
| TIPOS DE DOCUMENTOS | |
| Artículos | 408 |
| Reseña de libros | 98 |
| Artículo biográfico (*biographical-item*) | 1 |
| Material editorial (*editorial material*) | 1 |
| Reseñas (*reviews*) | 6 |
| CONTENIDOS DE LOS DOCUMENTOS | |
| Palabras-clave (*Keywords Plus*) | 129 |
| Palabras-clave de autor (*Author's Keywords*) | 1701 |
| AUTORES | |
| Autores | 617 |
| Autores de documentos de un sólo autor | 233 |
| COLABORACIÓN DE AUTORES | |
| Documentos de un sólo autor (*single-autored docs*) | 323 |
| Co-autores por documento % | 1.56 |
| Co-autoría internacional % | 3.3 |

Tabla de elaboración propia.

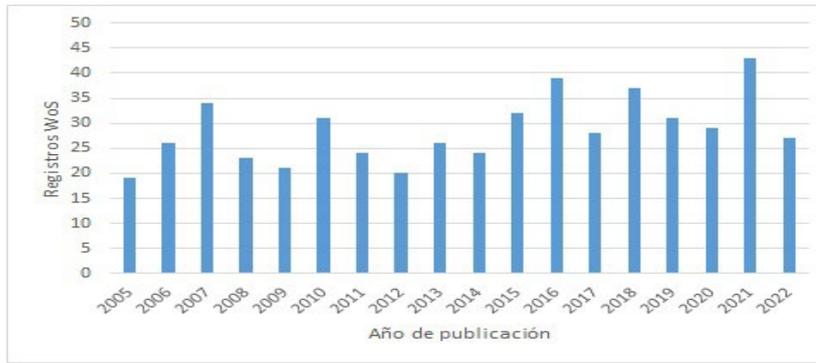

(figura.2) Número de registros en Web of Science (WoS) por año de publicación. Fuente: Elaboración propia.

Se obtuvieron un total de 1701 palabras-clave de autor, se mostraron las 20 palabras-clave de autor más frecuentes (Figura 3) y su dinámica de crecimiento por año (Figura 4). La visualización de las palabras-clave de autor en un mapa de árbol (o *Treemap*) mostraron los datos de forma jerárquica como un conjunto de rectángulos anidados (Figura 5). En el mapa de árbol, cada término ocupó un rectángulo, a través de figuras anidadas, las unas dentro de las otras, donde el tamaño de cada figura representó un valor métrico (por lo tanto, una caja más grande significó más ocurrencias): *Spain* (14%), *Archives* (7%), *Internet* (6%), *Spanish-Civil-War* (6%), *Documentation* (5%), *Libraries* (5%), *Photography* (5%), *Public libraries* (5%) o *Scientific production* (5%). A continuación, se realizó la etapa de pre-procesamiento para normalizar las palabras-clave seleccionadas.

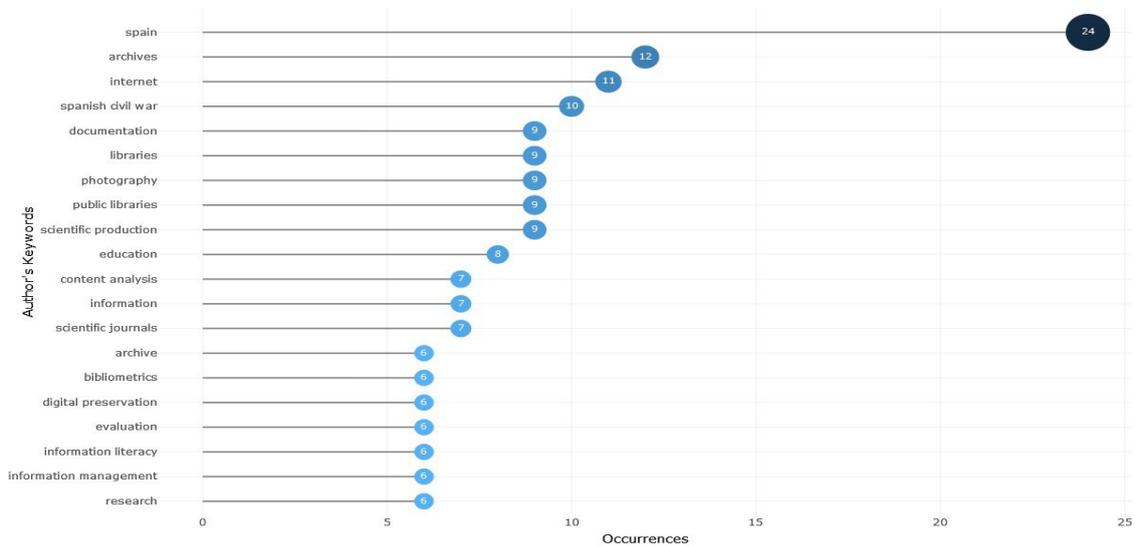

(figura.3) Las 20 palabras-clave de autor más frecuentes. Fuente: Elaboración propia..

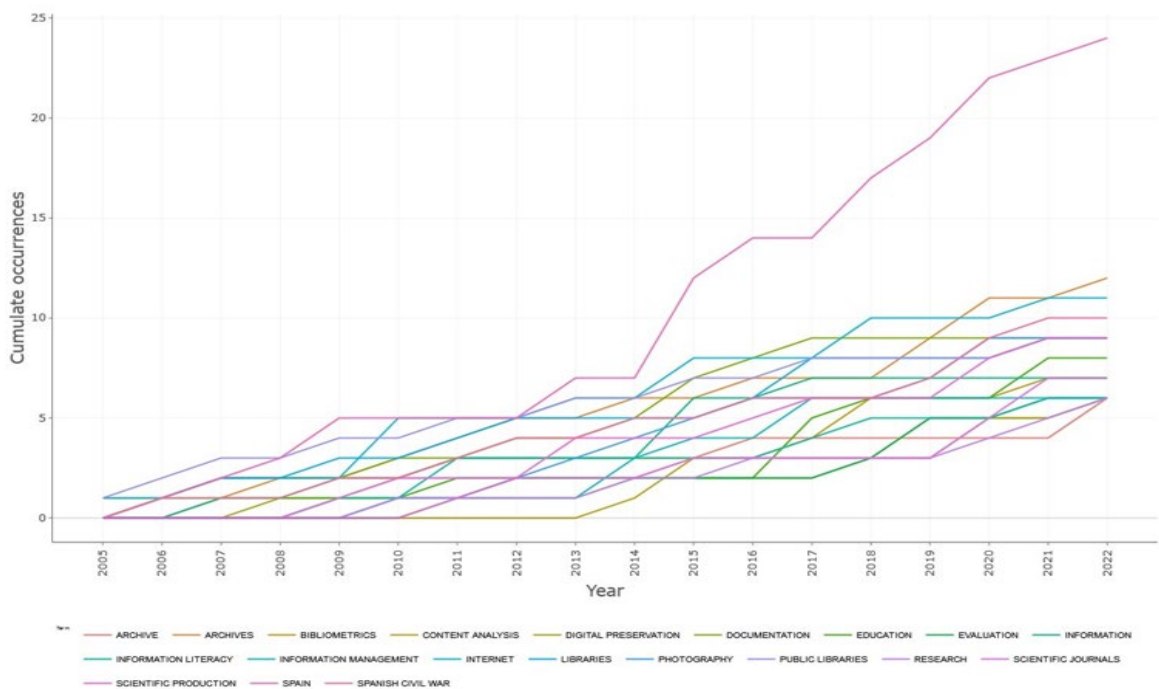

(figura.4) Dinámica de crecimiento de las 20 palabras-clave de autor más frecuentes. Fuente: elaboración propia.

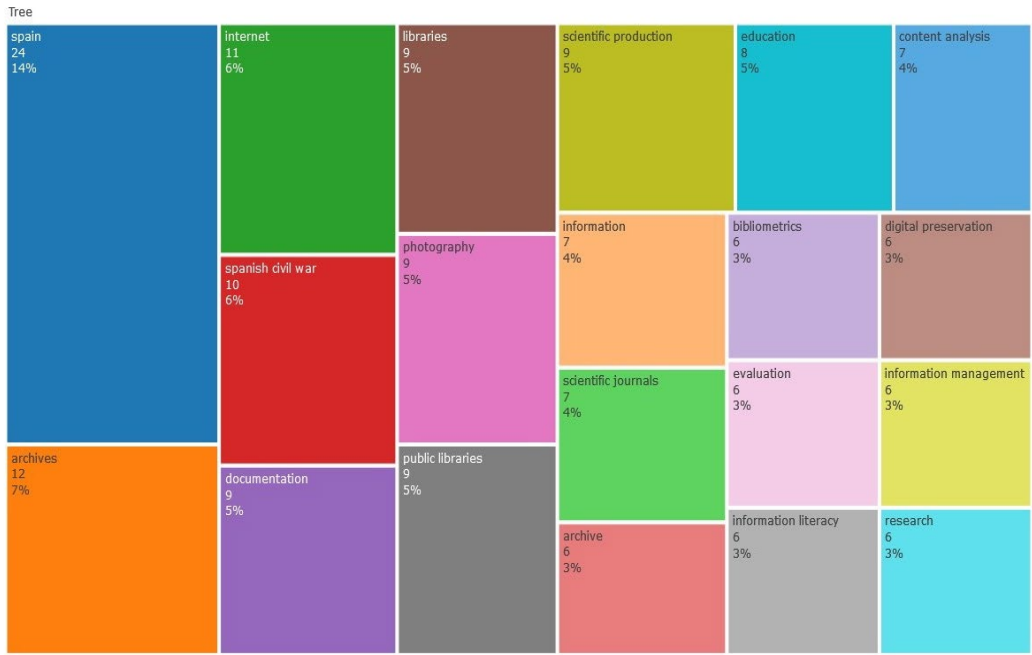

(figura.5) Mapa de árbol (o Treemap) de las 20 palabras-clave de autor más frecuentes. Fuente: elaboración propia.

Después de realizar la etapa de pre-procesamiento, en la que se estableció la división temporal de los datos en tres subperiodos: 2005-2010, 2011-2016 y 2017-2022. Se obtuvieron un total de 37 redes bibliométricas de palabras-clave (Figuras 6). Las redes bibliométricas se distribuyeron en la forma de mapas científicos (o diagramas estratégicos) según sus índices de centralidad y densidad. Por último, se creó el mapa de la evolución de los ciclos de vida de las áreas temáticas.

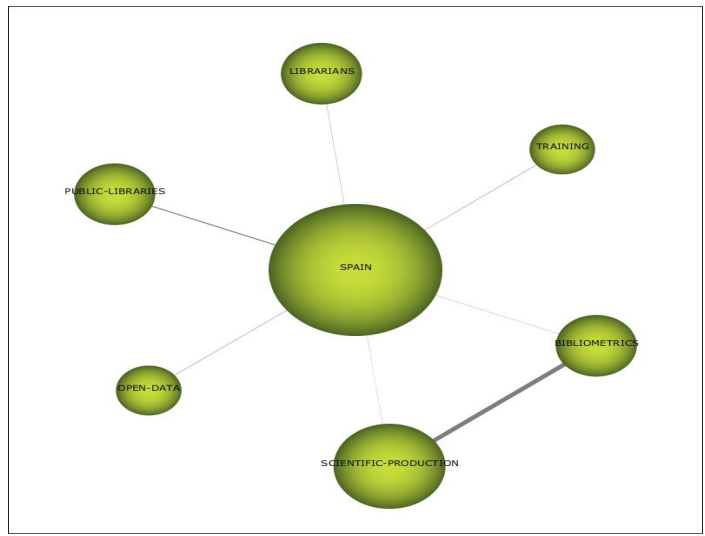

(figura. 6) Ejemplo de red bibliométrica de palabras-clave. Fuente: elaboración propia.

### 3.1. Mapa científico del periodo 2005-2010

De acuerdo con el diagrama estratégico del primer periodo 2005-2010 (Figura 7), se observó que la estructura del campo de investigación se distribuyó en torno a 8 redes temáticas (Tabla 2), clasificadas según su ubicación en los cuatro cuadrantes. Temas motores y centrales (sólidamente conectados a otros agregados, con un alto grado de desarrollo y de integración): *Internet* y *Book*. Temas básicos y transversales, *Spain* y *Students*. Temas especializados y bien desarrollados (por eso tienen una alta densidad) pero con enlaces externos débiles por eso se clasificaron como periféricos: *Bibliometric-Study, Old-Book* y *Information-Services*. Temas marginales y periféricos, con baja centralidad y densidad: *Archives.*

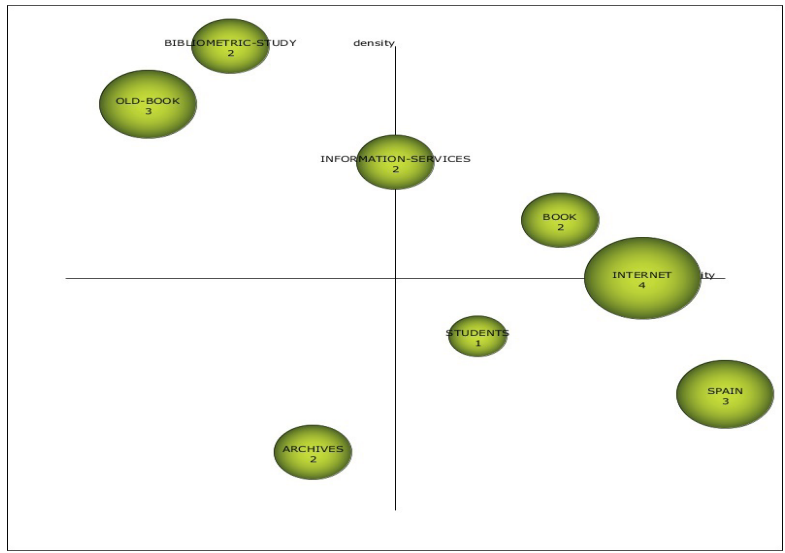

(figura.7) Mapa científico periodo 2005-2010. Fuente: Elaboración propia.

Tabla 2 Índices de centralidad y densidad de los temas (2005-2010).

| CUADRANTE (C) | TEMAS | NÚMERO DE DOCUMENTOS | CENTRALIDAD | DENSIDAD |
|---|---|---|---|---|
| Temas motores (C1) | Internet | 4 | 4.83 | 16.67 |
| | Book | 2 | 2.5 | 16.67 |
| Temas básicos (C2) | Spain | 3 | 7.75 | 6.25 |
| | Students | 1 | 0 | 12.5 |
| Temas especializados (C3) | Bibliometric-Study | 2 | 0 | 100 |
| | Old-Book | 3 | 0 | 62.25 |
| | Information-Services | 2 | 0 | 50 |
| Temas emergentes o marginals (C4) | Archives | 2 | 0 | 5.56 |

Tabla de elaboración propia.

### 3.2. El mapa científico del periodo 2011-2016

El diagrama estratégico del periodo 2011-2016 (Figura 8) mostró que la estructura científica giró en torno a 11 grandes redes temáticas (Tabla 3), clasificadas en los cuatro cuadrantes. Temas motores y centrales: Documentation, Spain y Quality. Temas básicos y transversales, Photography, Bibliography y Communication. Temas especializados, coherentes, bien desarrollados y periféricos: France, Scientific-Journals y Document-Management. Temas marginales, con baja centralidad y densidad: Libraries y Information-Management.

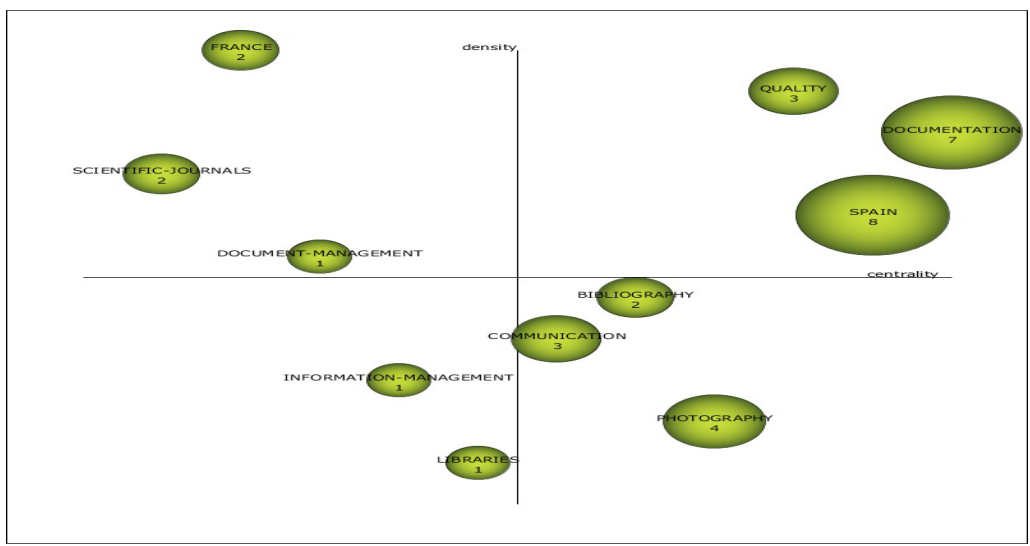

(figura.8) Mapa científico periodo 2011-2016. Fuente: Elaboración propia.

Tabla 3 Índices de centralidad y densidad de los temas (2011-2016).

| CUADRANTE (C) | TEMAS | NÚMERO DE DOCUMENTOS | CENTRALIDAD | DENSIDAD |
|---|---|---|---|---|
| Temas motores (C1) | Documentación | 7 | 18.85 | 18.96 |
| | Spain | 8 | 12.72 | 13.65 |
| | Quality | 3 | 9.72 | 27.78 |
| Temas básicos (C2) | Photography | 4 | 8.33 | 7.33 |
| | Bibliography | 2 | 6.94 | 11.11 |
| | Communication | 3 | 5.28 | 11.11 |
| Temas especializados (C3) | France | 2 | 0 | 36.11 |
| | Scientif-Journals | 2 | 0 | 16.67 |
| | Document-management | 1 | 0 | 12.5 |
| Temas emergentes o marginals (C4) | Libraries | 1 | 3.96 | 4.17 |
| | Information-Management | 1 | 1.94 | 8.33 |

Tabla de elaboración propia.

### 3.3. El mapa científico del periodo 2017-2022

En el diagrama estratégico del periodo 2017-2022 (Figura 9), la estructura científica se distribuyó en torno a 18 redes temáticas, clasificadas en los cuatro cuadrantes (Tabla 4). Temas motores y centrales: *Evaluation, Spain, Education, Bibliometrics, Internet* y *Archives*. Temas básicos y transversales: *University-Libraries, Knowledge-Management, Documentary-Analysis y Digital-Libraries*. Temas especializados, coherentes, bien desarrollados y periféricos: *Spanish-Civil-War, Letters, Thematic-Categorization y Women-Phographers*. Temas emergentes o marginales, con baja centralidad y densidad: *Covid-19, Indulgence, Digital-Preservation* y *Archival-Science.*

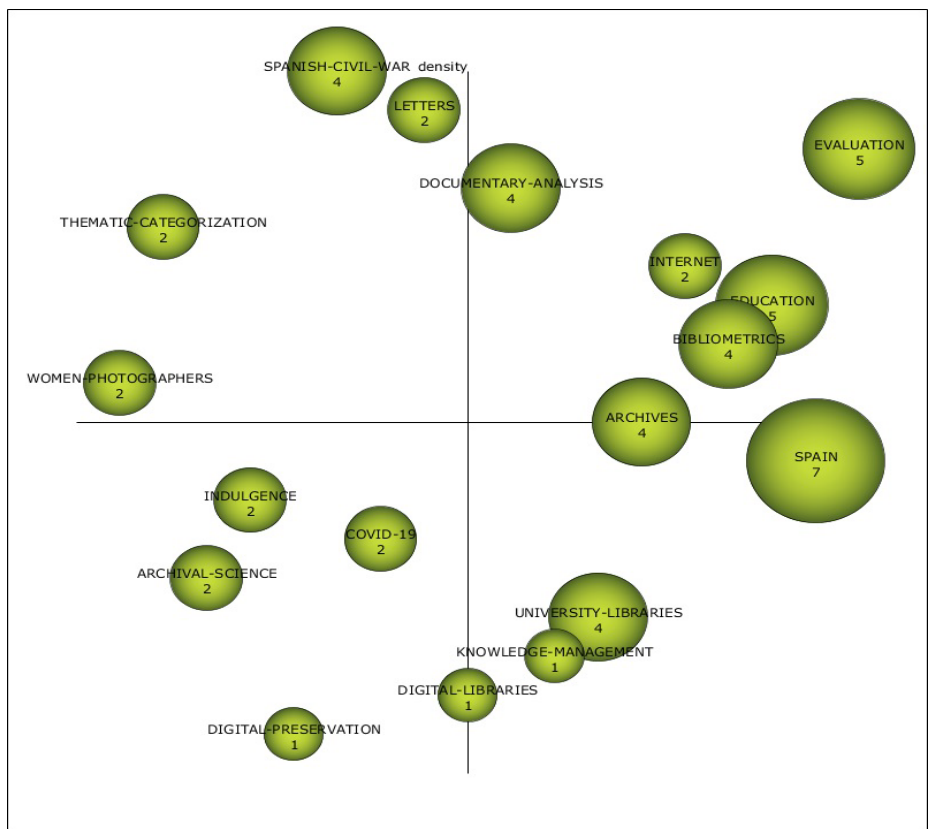

(figura.9) Mapa científico periodo 2017-2022. Elaboración propia.

Tabla 4 Índices de centralidad y densidad de los temas (2011-2016).

| CUADRANTE (C) | TEMAS | NÚMERO DE DOCUMENTOS | CENTRALIDAD | DENSIDAD |
|---|---|---|---|---|
| Temas motores (C1) | Evaluation | 5 | 40.42 | 45 |
| | Spain | 7 | 39.61 | 16 |
| | Education | 5 | 36.62 | 26.56 |
| | Bibliometrics | 4 | 19.33 | 16.84 |
| | Internet | 2 | 17.5 | 26.85 |
| | Archives | 4 | 16.86 | 16.37 |
| Temas básicos (C2) | University-Libraries | 4 | 14.94 | 9.72 |
| | Knowledge-Management | 1 | 9.36 | 8.33 |
| | Documentary-Analysis | 4 | 8.33 | 34.17 |
| | Digital-Libraries | 1 | 6.67 | 8.33 |
| Temas especializados (C3) | Spanish-Civil-War | 4 | 4.17 | 90.97 |
| | Letters | 2 | 5 | 50 |
| | Thematic-Categorization | 2 | 0.25 | 29.17 |
| | Women-Photographers | 2 | 0 | 16.67 |
| Temas emergentes o marginals (C4) | Covid-19 | 1 | 4.17 | 11.11 |
| | Indulgence | 1 | 2.5 | 13.89 |
| | Digital-Preservation | 1 | 2.5 | 6.25 |
| | Archival-Science | 2 | 1.25 | 11.11 |

Tabla de elaboración propia.

### 3.4. Mapa de evolución de las áreas temáticas

El procedimiento para analizar la evolución conceptual y temática del campo científico, tratado en la revista, se realizó a través de la evolución de las palabras-clave y de las 37 grandes áreas temáticas detectadas.

En primer lugar, se midió el grado de solapamiento entre las palabras-clave utilizadas en la revista a lo largo del tiempo (Figura 10). En la evolución de las palabras-clave, los círculos representaron cada periodo de tiempo, el número dentro de ellos representa el total de palabras-clave del periodo correspondiente. Las flechas entre dos periodos consecutivos representaron el número de palabras-clave compartidas por ambos periodos, entre paréntesis se mostró el índice de estabilidad (o grado de solapamiento). Las flechas salientes representaron las palabras-clave que no se utilizaron por el siguiente periodo de tiempo (esto es, el número de palabras-clave que no tuvieron continuidad temporal en el periodo de tiempo inmediatamente posterior). Las flechas entrantes representan el número de palabras-clave nuevas introducidas en el periodo correspondiente. Los resultados fueron los siguientes:

− Primer periodo 2005-2010: se utilizaron 482 palabras-clave, de ellas 60 permanecieron en el intervalo siguiente (2011-2016) y las restantes 422 no se volvieron a utilizar, considerándose términos efímeros. El índice de similitud entre el primer y segundo periodo fue de 0,12.
− Categoría 2: Segundo periodo 2011-2016: se utilizaron 608 palabras-clave, se incorporaron 548 palabras nuevas. En el intervalo siguiente permanecieron 95 y 513 no se volvieron a utilizar, el índice de estabilidad entre el segundo y tercer periodo aumentó y fue de 0,16.
− Tercer periodo 2017-2022: fue el que contó con un número mayor de palabras, 762. Incorporó 667 nuevas palabras-clave.

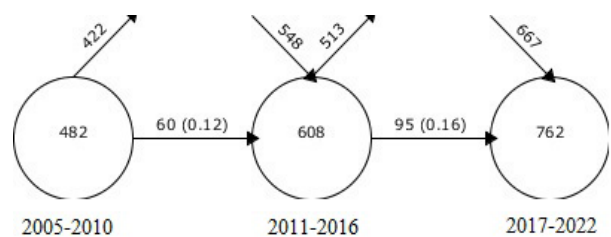

(figura.10) Evolución de palabras-clave entre periodos. Elaboración propia.

En segundo lugar, el mapa de evolución resultante (Figura 11) nos permitió mostrar cómo progresó la estructura científica a lo largo del tiempo a través de las áreas temáticas que tuvieron mayor interés. En el mapa, cada columna correspondió al periodo establecido (2005-2010, 2011-2016 y 2017-2022), se tuvieron en cuenta diferentes aspectos:

- El volumen de las esferas fue proporcional al número de documentos que albergó cada tema.
- Las líneas continuas representaron los nexos temáticos.
- Las líneas discontinuas de puntos reflejaron los temas vinculados que compartieron palabras-clave diferentes al nombre, o etiqueta, de los temas.
- El grosor de las líneas entre dos temas fue proporcional al índice de inclusión.

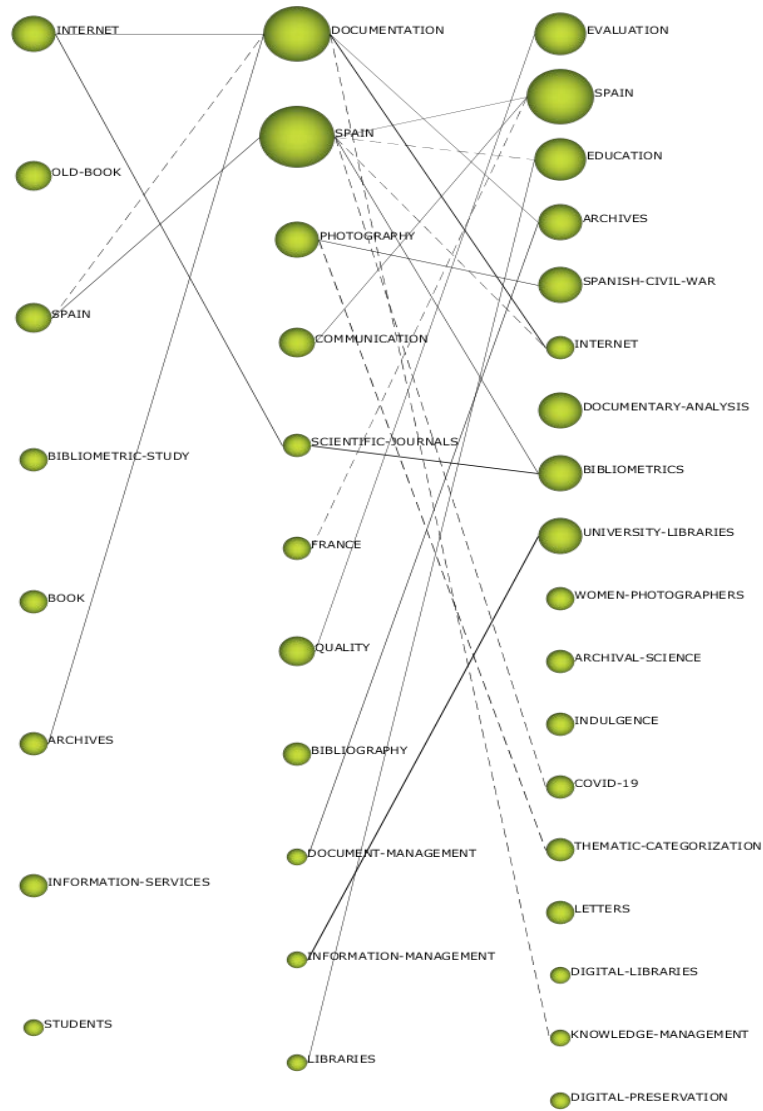

(figura.11) Mapa de evolución de las áreas temáticas. Elaboración propia.

## 4. Discusión

En el periodo 2005-2010, los clústeres temáticos se distribuyeron fundamentalmente en la segunda bisectriz (cuadrante 2 - cuadrante 3). La organización del campo de investigación se clasificó en la categoría 2. Pocos temas se ubicaron en las posiciones centrales (*Book, Internet* y *Spain*). Presentó agregados en el cuadrante 3, que representaron redes de palabras-clave periféricas, pero bien desarrolladas, que se identificaron con temas especializados (*Bibliometric-Study* y *Old-Book*). La estructura del campo de investigación estaba en una etapa inicial, con muy pocos temas centrales y básicos.

En el periodo 2011-2016, las redes temáticas se distribuyeron en los diferentes cuadrantes, pudiéndose considerar una incipiente organización sólida, incluida en la categoría 3. Sin embargo, aunque se presentaron familias de todos los posibles agregados (estabilizados, emergentes o en vías de desarrollo), todavía no se pudo definir como una estructura consolidada. Unos pocos agregados ocuparon posiciones centrales, representando temas motores y bien desarrollados (*Documentation, Spain, Quality, Photography, Bibliography* y *Communication*), también se mostraron pocos temas periféricos y especializados. En esta fase temporal intermedia, la estructura del campo de investigación estaba en vías de estructuración.

En el periodo 2017-2022, las redes temáticas se distribuyeron en los diferentes cuadrantes, pudiéndose considerar una estructura consolidada, clasificada de forma inequívoca en la categoría 3. La distribución de los agregados estuvo en los cuatro cuadrantes, se encontraron todo tipo de familia de temas (centrales, especializados, con distintos grados de desarrollo, transversales, emergentes o en vías de desaparición). Destacaron diversos temas motores y básicos relacionados con los estudios métricos, el análisis documental y la archivística (*Evaluation, Spain, Education, Bibliometrics, Internet, Archives*, *Documentary-Analysis*). También aparecieron temas especializados en investigación histórica (*Spanish-Civil-War, Letters, Indulgence*) o temas emergentes relacionados con la pandemia (*Covid-19*). La estructura del campo de investigación se calificó como consolidada, con una dinámica compleja, heterogénea y valiosa.

En cuanto a la evolución de las palabras-clave de los periodos analizados (según su grado de solapamiento, continuidad o discontinuidad) se constató que el conjunto de descriptores no se mantuvo constante en los diferentes intervalos. Se comprobó que la terminología se fue transformando, asignando diferentes palabras-clave para representar el contenido de los documentos. El índice de estabilidad se mantuvo relativamente constante a lo largo del tiempo, lo que nos indicó que la revista ha ido afianzando un restringido vocabulario para describir a los documentos publicados. Por otro lado, el número de palabras-clave que no se utilizaron en los periodos consecutivos fue muy elevado, implicando un gran número de palabras-clave transversales y efímeras, que nunca más se volvieron a utilizar. Con lo anterior, se constató que las palabras-clave se han ido incrementando y renovando significativamente a lo largo de los periodos analizados, surgiendo términos nuevos y desapareciendo otros.

En relación a las áreas temáticas, el mapa nos permitió explicar la evolución de las áreas temáticas que centraron el interés de la comunidad científica:

a) Entre el primer periodo (2005-2010) y el segundo (2011-2016) no hubo muchos nexos conceptuales entre las áreas temáticas, se detectó una falta de cohesión e interrupciones (o huecos) en la evolución.
b) En el tercer periodo (2017-2022) encontramos que las áreas temáticas se originaron a partir de áreas asociadas con el periodo anterior, esto constató que el campo de investigación presentó en los dos últimos periodos una gran continuidad, coherencia y desarrollo.
c) El área temática *Spain* estuvo presente en todos los periodos estudiados y mostró líneas continuas y discontinuas con otras áreas temáticas, esto demostró que la revista ha publicado de forma continua trabajos relacionados con distintas áres temáticas relacionadas con este ámbito geográfico.
d) La mayor parte de las líneas continuas unieron áreas temáticas con una sólida evolución, tales como: *Internet*, *Archives*, *Documentation*, *Photography*, *Communication*, *Scientific-Journal*, *Quality*, *Information-Management*, *Evaluation*, *Education*, *Spanish-Civil-War*, *Bibliometrics*, *University-Libraries*.
e) Otras áreas temáticas, como *Covid-19*, *Archival-Science*, *Digital-Preservation* o *Indulgence*, no se vincularon a ninguna área temática previa, considerándose áreas emergentes o en vías de desaparición.

## 5. Conclusiones

Sed La metodología utilizada nos permitió representar la morfología del campo de investigación de la publicación RGID y estudiar las trayectorias privilegiadas de algunas áreas temáticas durante los 18 años analizados, según los datos obtenidos de WoS. El método empleado consistió en transformaron datos abstractos y relaciones complejas en visualizaciones comprensibles, que nos permitieron extraer conocimiento. Las principales conclusiones fueron las siguientes:

– Se obtuvieron un total de 1701 palabras-clave de autor, entre las más frecuentes destacaron: *Spain*, *Archives*, *Internet*, *Spanish-Civil-War*, *Documentation*, *Libraries*, *Photography*, *Public libraries* o *Scientific production*.
– En el periodo 2005-2010, unos pocos grupos temáticos (*Internet* y *Book*) se ubicaron en el cuadrante de los temas centrales. La distribución de los temas en los cuadrantes mostró un campo en su fase inicial, con muy pocos grupos temáticos motores y básicos
– En el periodo 2011-2016, sólo algunos agregados ocuparon posiciones centrales, representando temas motores y básicos (*Documentation, Spain, Quality, Photography, Bibliography* y *Communication*). La distribución de los temas en los cuadrantes, indicó un campo en vías de estructuración, con temas motores, centrales y especializados, pero poco numerosos.
– En el periodo 2017-2022, destacaron diversos temas motores, básicos y especializados, relacionados con la evaluación, los estudios métricos, el análisis documental, la archivística (*Evaluation, Spain, Education, Bibliometrics, Internet, Archives*, *Documentary-Analysis*). La distribución de la amplia familia de temas en todos los cuadrantes, reveló un campo consolidado, con una estructura compleja y rica.
– En cuanto a la evolución de las palabras-clave de autor, se constató que el conjunto de descriptores, aplicado para la indización de documentos, no se mantuvo constante en los diferentes intervalos temporales. Los resultados revelaron que la terminología utilizada progresó de forma continua en los periodos analizados, surgiendo muchos términos nuevos y desapareciendo otros.
– En cuanto a la evolución de las áreas temáticas, a través del análisis del mapa longitudinal, se mostraron los patrones temáticos en los diferentes periodos. El área temática *Spain* estuvo presente en todos los periodos estudiados. Diversas áreas temáticas presentaron una sólida evolución (fundamentalmente entre los dos últimos periodos 2011-2016 y 2017-2022), tales como: *Internet*, *Archives*, *Documentation*, *Photography*, *Communication*, *Scientific-Journal*, *Quality*, *Information-Management*, *Evaluation*, *Education*, *Spanish-Civil-War*, *Bibliometrics*, *University-Libraries*.
– La dinámica de las áreas temáticas corroboró los resultados obtenidos en los diagramas estratégicos, se calificaron las etapas del campo de investigación en: inicio o creación (2005-2010), vías de desarrollo (2011-2016) y consolidación (2017-2022).

El estudio retrospectivo realizado no se dirigió a la aceptación automática de los resultados para la toma de decisiones en distintos ámbitos, sino a ofrecer una impresión visual para que los editores y autores puedan

C. Gálvez 31.1 (2024): 127-140                                                                                                                    139identificar los focos de investigación más relevantes, el ciclo de vida de algunas áreas temáticas o mostrar la política editorial que ha seguido RGID a lo largo del tiempo. No obstante, se debe tener en cuenta que cada instrumento de análisis tiene sus propias limitaciones, en este caso la construcción de los mapas científicos ha dependido de la calidad de las indizaciones realizadas por los propios autores.

## 6. Referencias bibliográficas

Abadal, Ernet. [2017]. *Revistas científicas. Situación actual y retos de futuro*. Barcelona: Edicions Universitat Barcelona.

Abadal, Ernet. [2018]. "¿Cómo han cambiado BiD y las revistas españolasde documentación en los últimos veinte años?" *BiD: textos universitaris de biblioteconomia i documentació*, 40: 1-10. DOI: https://dx.doi.org/10.1344/BiD2018.40.11

Aria, Massimo; Cuccurullo Corrado [2017]. "Bibliometrix: An R-tool for comprehensive science mapping analysis". *Journal of Informetrics,* 11 (4): 959-75. DOI: https://doi.org/10.1016/j.joi.2017.08.007

Bonnevie, Ellen [2003]. "A multifaceted portrait of a library and information science journal: The case of the Journal of Information Science". *Journal of Information Science*, 29 (1): 11-24. DOI: 10.1177/0165551103762202032

Bordons, María; Zulueta, Mª Ángeles [1999]. "Evaluación de la actividad científica a través de indicadores bibliométricos". *Revista Española de Cardiología*, 52 (10): 790-800.

Börner, Katy; Chen, Chen.; Boyack, K. Woyack [2003]. "Visualizing knowledge domains". *Annual Review of Information Science and Technology*, 37, 179-255. DOI: https://doi.org/10.1002/aris.1440370106

Callon, M.; Courtial, J. P.; Laville, F. [1991]. "Co-word analysis as a tool for describing the network of interactions between basic and technological research: The case of polymer chemistry". *Scientometrics*, 22: 155-205. DOI: https://doi.org/10.1007/BF02019280

Callon, Michel; Rip, Arie; Law, John [1986]. *Mapping the Dynamics of Science and Technology*. London: The Macmillan Press Ltd.

Cano, V. [1999]. "Bibliometric overview of library and information science research in Spain". *Journal of the American Society for Information Science*, 50 (8): 675-80. DOI: https://doi.org/10.1002/(SICI)1097-4571(1999)50:8<675::AID-ASI5>3.0.CO;2-B

Cascón Katchadourian, Jesús; Moral-Munoz, José A; Liao, Huchang; Cobo, Manuel J. [2020]. "Análisis bibliométrico de la Revista Española de Documentación Científica". *Revista Española de Documentación Científica*, 43. DOI: https://doi.org/10.3989/redc.2020.3.1690

Cobo, Manuel Jesús [2011]. *SciMat: Herramienta software para el análisis de la evolución del conocimiento científico: Propuesta de una metodología de evaluación*. Tesis doctoral, Universidad de Granada.

Cobo, Manuel Jesús; López-Herrera, A. G.; Herrera-Viedma, E.; Herrera, F. [2011]. "An approach for detecting, quantifying, and visualizing the evolution of a research field: A practical application to the fuzzy sets theory field". *Journal of Informetrics*, 5 (1): 146-66. DOI: https://doi.org/10.1016/j.joi.2010.10.002

Cobo, M. J.; López-Herrera, A. G.; Herrera-Viedma, E.; Herrera, F. [2012]. "SciMAT: A new science mapping analysis software tool". *Journal of the American Society for Information Science and Technology*, 63 (8): 1609-30. DOI: https://doi.org/10.1002/asi.22688

Coulter, Neal; Monarch, Ira; Konda, Suresh [1998]. "Software engineering as seen through its research literature: A study in co-word analysis". *Journal of the American Society for Information Science*, 49 (13): 1206-23. DOI: https://doi.org/10.1002/(SICI)1097-4571(1998)49:13<1206::AID-ASI7>3.0.CO;2-F

DeHart, Florence E. [1992]. "Monographic references and information science journal literature". *Information Processing & Management*, 28 (5): 629-35.

Giménez-Toledo, Elea; Román-Román, Adelaida [2000]. "Evaluación de revistas científicas: análisis comparativo de dos modelos y su aplicación a cinco revistas españolas de biblioteconomía y documentación". *Interciencia*, 25 (5): 234-41. DOI: http://www.redalyc.org/articulo.oa?id=33904703

Glänzel, Wolfgang; Moed, Henk F. [2013]. "Opinion paper: thoughts and facts on bibliometric indicators]. *Scientometrics*, 96 (1): 381-94. DOI: https://doi.org/10.1007/s11192-012-0898-z

Guallar, Javier; Ferran-Ferrer, Nuria; Abadal, Ernest; Server, Adán [2017]. "Revistas científicas españolas de información y documentación: análisis temático y metodológico". *Profesional de la Información*, 26 (5): 947-60. DOI: https://doi.org/10.3145/epi.2017.sep.16

Guallar, Javier; López-Robles, José-Ricardo; Abadal, Ernest; Gamboa-Rosales, Nadia-Karina; Cobo, Manuel Jesús [2020]. "Revistas españolas de Documentación en Web of Science: análisis bibliométrico y evolución temática de 2015 a 2019". *Profesional de la Información*, 29 (6). DOI: https://doi.org/10.3145/epi.2020.nov.06

Jiménez-Hidalgo, Sonia [2007]. "Análisis de la autoría en la Revista Española de Documentación Científica (1997-2005)". *Revista Española de Documentación Científica,* 30 (3): 305-22. DOI: https://doi.org/10.3989/redc.2007.v30.i3.387

Koehler, Wallace [2001]. "Information science as "Little Science": The implications of a bibliometric analysis of the Journal of the American Society for Information Science". *Scientometrics*, 51 (1): 117-32. DOI: 10.1023/A:1010516712215

Leydesdorff, Loet; Welbers, Kasper [2011]. "The semantic mapping of words and co-words in contexts". *Journal of Informetrics*, 5: 469-75. DOI: https://doi.org/10.1016/j.joi.2011.01.008

Lipetz, Ben Ami [1999]. "Aspects of JASIS Authorship through Five Decades". *Journal of the American Society for Information Science and Technology,* 50 (11): 994-1003. DOI: https://doi.org/10.1002/(SICI)1097-4571(1999)50:11<994::AID-ASI5>3.0.CO;2-U